\documentstyle[aps,epsfig,prb]{revtex}
\begin{document}
\draft
\title{Impurity-induced Ferromagnetism in Doped Triplet Excitonic Insulator}
\author{Takashi Ichinomiya}
\address{Department of Material Physics, Faculty of Engineering Science, Osaka University,
Toyonaka 560-8531,Japan}
\date{\today}
\maketitle
\begin{abstract}
The theory of impurities in excitonic insulator is investigated in the
 light of the recent experiments on hexaborides.  First, we
 study the bound state around the impurity and find that the bound
 state emerges when $\mbox{Re}\Delta$ is positive. Second, we study the
 continuum state using Abrikosov-Gor'kov's approach. We find that the
 energy gap is reduced strongly when
 $\mbox{Im}\Delta=0$. Finally, we solve
 Bogoliubov-de Gennes equations for excitonic insulator
 numerically. We get the  results  consistent with the analytic ones.
 We also find that incomplete ferromagnetism appears in doped triplet
 excitonic insulator with impurity. We make a short qualitative
 discussion on the ferromagnetism of doped hexaborides using our result. 
\end{abstract}
\pacs{71.10.Ca, 71.35.-y, 75.10.Lp}
\section{Introduction}
\label{sec:introduction}
The discovery of weak ferromagnetism in La-doped CaB$_6$ is  one
of the greatest surprises in the recent study of ferromagnetism 
\cite{Young1}. This material shows ferromagnetism with high Curie
temperature about $T_{\mbox{c}}\sim 600K$ and with small magnetic moment.  
 Zhitomirsky {\it et al.}\cite{Rice1} claimed that the theory of the
 excitonic ferromagnetism, which was originally proposed by Volkov {\it
 et al.}\cite{Volkov1}, can account for this curious ferromagnetism. 
Their proposal has brought  renewed interests in excitonic insulator,
 which was vigorously studied theoretically in 1960s.

 After the proposal by Zhitomirsky {\it et al.}, some authors tried to improve
 their theory. Balents and Varma\cite{Varma1} and Barzykin and Gor'kov\cite{Gorkov1} independently discussed that, to explain this
 ferromagnetism, the approximation of excitonic insulator should be
 improved by considering the formation of  superstructure or long-range
 Coulomb  interaction.

 The effect of impurities will also be important for the  excitonic
 insulator. In this paper, we study the effect of impurities in excitonic
 insulators.  

 The destruction of excitonic insulator by the impurity was
 discussed by Zittartz\cite{Zittartz}. He showed that the intraband
 scattering by impurity reduces the energy gap just as the magnetic
 impurity in superconductor. In this paper we study the interband
 scattering by impurity. We show that the interband scattering cause the
 formation of the bound states and the reduction of the energy gap.
 We also show that this interband scattering breaks the $U(1)$ symmetry of
 hamiltonian. This is due to the fact that
 the phase of the order parameter is determined by the difference of
 phase between conduction and valence bands. The reduction of
 the energy gap and the energy of the the bound state is controlled by
 the phase of order parameter. The phase dependence of the energy gap and
 the bound state brings the ferromagnetic ordering of doped electrons 
 in triplet excitonic insulator, because the phase of order parameter
 differs by $\pi$ between $\Delta_{\uparrow}$ and $\Delta_{\downarrow}$. 

 This paper is constructed as follows. First, in the next section we
 introduce the Hamiltonian and show  the breaking of $U(1)$ symmetry 
 by the interband scattering. In sec. \ref{sec:bound-state}  and
 \ref{sec:gap-reduction} we
 investigate the effect of impurities on the bound and continuum
 states. It is shown that the bound states emerge only when 
 $\mbox{Re}\Delta$ is positive and that the binding energy depends on
 the phase of the order parameter. We also
 show that the in the continuum states  the energy gap is strongly suppressed 
 when $\mbox{Re}\Delta=0$. In sec. \ref{sec:bdg},  we solve
 Bogoliubov-de Gennes equation for excitonic insulator numerically. The
 results support the analytic discussion above. We also find that, when
 $\phi$ is close to $\pi/2$, magnetic moment of doped electrons is reduced. 
 In the last section for summary and  discussion, we  discuss about the 
 application of our results to the ferromagnetism in  hexaborides. 

\section{Breaking of $U(1)$ symmetry by Interband Scattering}
\label{sec:breaking-symmetry}

First we study the following  hamiltonian 
\begin{equation}
 H_{0} = \sum_{k,\sigma}( \epsilon^a_{k}a^{\dagger}_{k,\sigma}a_{k,\sigma}+ \epsilon^b_{k}b^{\dagger}_{k,\sigma}b_{k,\sigma}) + \frac{V}{\Omega}\sum_{k,k^{\prime},q,\sigma,\sigma^{\prime}} a^{\dagger}_{k,\sigma}b^{\dagger}_{q-k,\sigma^{\prime}}b_{q-k^{\prime},\sigma^{\prime}}a_{k^{\prime},\sigma} + \frac{U}{\Omega} \sum_{k,k^{\prime},q,\sigma,\sigma^{\prime}} a^{\dagger}_{k,\sigma}b^{\dagger}_{q-k,\sigma^{\prime}}a_{q-k^{\prime},\sigma^{\prime}}b_{k^{\prime},\sigma}\label{cleanHamiltonian}.
\end{equation} 
 Here $a_{k,\sigma}$ and $b_{k,\sigma}$ are annihilation operators of
 electrons in the lower
 and upper band, $\epsilon^a_{k}$ and $\epsilon^b_{k}$ are kinetic
 energy of electrons, and $V$ and $U$ are exchange  and direct interaction
 between two bands. The nesting of Fermi surfaces between two the electron
 bands brings the formation of excitonic order parameter\cite{Rice2}.

 We introduce the interband scattering  by impurity  $H_{\mbox{imp}}$ as
\begin{equation}
 H_{\mbox{imp}}=\sum V_{\mbox{imp}}a^{\dagger}_{i\sigma}b_{i\sigma} + c.c. 
\label{impurityHamiltonian}
\end{equation} 

 This interband scattering is important because this term breaks the $U(1)$
 symmetry of the
 system. The  hamiltonian(\ref{cleanHamiltonian}) for the clean system
 is invariant under the
 $U(1)\times U(1)$ transformation, $a_i \rightarrow
 a_i e^{i \phi_1}$ and $b_i \rightarrow b_ie^{i \phi_2}$. However,
 the interband scattering term reduces this symmetry to
 the diagonal $U(1)$  symmetry because the  $U(1) \times U(1)$ transformation
 change the impurity term as  
 $V_{\mbox{imp}}a^{\dagger}_{i\sigma} b_{i\sigma} \rightarrow
 V_{\mbox{imp}}a^{\dagger}_{i\sigma}b_{i\sigma} e^{i(\phi_2 - \phi_1)}$,
 unless $\phi_1 \neq \phi_2$. This means that to discuss the impurity
 effect we must take  the phase of excitonic order parameter, which
 corresponds to $\langle a^{\dagger} b \rangle$, into account.

 Here we note that in real materials this $U(1)$ symmetry is generally broken even without
 impurities due to  other interactions which are
 not included in the Hamiltonian(\ref{cleanHamiltonian}). The interband
 pair scattering term 
 $V^{\prime}a^{\dagger}_{k\sigma}a^{\dagger}_{k^{\prime}-\sigma}b_{k^{\prime}-\sigma}b_{k\sigma}$ is one of such interactions. Therefore we suppose the
 phase $\phi$ is determined by such term. It is interesting to study the
 stable phase $\phi$ and discuss the phase dynamics, however, this is beyond
 the scope of this paper.  

 Volkov {\it et al.} noticed that localized  states with uncompensated
 spin appears  in the doped triplet excitonic
 insulator\cite{Volkov2}. They concluded that the bound states appears
 only when $\Delta$ is positive. Their result means that the phase of
 the  order parameter determines the existence of bound state. However, they
 assume that  $\Delta$ to be real, which is needed to be justified by further
 study. In this paper, we treat the order parameter as the complex variable. 
 Here we introduce the excitonic mean
  field order parameter 
 $\Delta_{\sigma\sigma^{\prime}}$ and write down the mean-field
 hamiltonian in pure excitonic system 
\begin{equation}
 H_{\mbox{MF}}=\sum_{\sigma,k} \epsilon^a_k a^{\dagger}_{k\sigma}a_{k\sigma} + \epsilon^b_{k}b^{\dagger}_{k\sigma}b_{k\sigma} - \sum_{\sigma,\sigma^{\prime},k} (\Delta_{\sigma\sigma^{\prime}}b^{\dagger}_{k\sigma^{\prime}}a_{k\sigma} + \Delta^{\dagger}_{\sigma\sigma^{\prime}}a^{\dagger}_{k\sigma^{\prime}}b_{k\sigma}) -\mu\sum_{\sigma,k}(a^{\dagger}_{k\sigma}a_{k\sigma}+b^{\dagger}_{k\sigma}b_{k\sigma}).\label{MFHamiltonian}
\end{equation} 
When we use the hamiltonian (\ref{cleanHamiltonian}), order parameter
$\Delta_{\sigma\sigma{\prime}}$ is defined as 
\begin{equation}
 {\bf \Delta}_{\sigma\sigma^{\prime}}=V\sum_{k}\langle a^{\dagger}_{k\sigma}b_{k\sigma^{\prime}}\rangle -U \sum_{k,\sigma{\prime\prime}}\delta_{\sigma\sigma^{\prime}}\langle a^{\dagger}_{k\sigma^{\prime\prime}}b_{k\sigma^{\prime\prime}}\rangle.
\label{op-equation-wavespace}
\end{equation}
This Hamiltonian is invariant under the transformation $a_i \rightarrow
 a_i e^{i \phi_1}$,  $b_i \rightarrow b_i e^{i \phi_2}$ and
 $\Delta_{\sigma\sigma^{\prime}} \rightarrow e^{i(\phi_2 -
 \phi_1)}$. It is convenient to assume  the order parameter
 $\Delta_{\sigma\sigma^{\prime}}$ to be real by choosing the appropriate
 gauge.
 In this  case, the impurity scattering term (\ref{impurityHamiltonian}) shows
 phase dependence, 
\begin{equation}
  H^{\prime}_{\mbox{imp}}=\sum V_{\mbox{imp}}e^{i \phi}a^{\dagger}_{i\sigma}b_{i\sigma} + c.c. \label{impurity2}
\end{equation}
 In the following two sections, we use Hamiltonian(\ref{MFHamiltonian})
 and (\ref{impurity2}).

\section{Bound State Formation at an impurity}
\label{sec:bound-state}

First, we study the bound state around the  impurity in excitonic
 insulator. We here consider the one-impurity problem. For simplicity, we assume the order parameter
 $\Delta_{\sigma\sigma^{\prime}}$
to be diagonal, $\Delta_{\sigma\sigma^{\prime}}= \Delta_\sigma
\delta_{\sigma\sigma^{\prime}}$. In the singlet excitonic insulator this
 assumption is correct. We note that in the triplet excitonic insulator
 we can justify this assumption  if the three component of triplet order
 parameter  $\Delta_i =
 \sum_{\alpha,\beta}(\sigma_i)_{\alpha\beta}\Delta_{\alpha\beta}$
 ($i=x, y, z$) are degenerate.
In this case, we
can treat up-spin and down-spin separately. 
Then from (\ref{MFHamiltonian}) the Green's function in the mean-field theory is
 given by
\begin{equation}
 G^0_{\sigma\sigma^{\prime}}(k,\omega)=\left(
\matrix{
\omega-\epsilon^a_k + \mu  &- \Delta_\sigma \cr
-\Delta^{*}_\sigma & \omega-\epsilon^b_k+\mu \cr
}
\right)^{-1}\delta_{\sigma\sigma^{\prime}} .
\end{equation}

 Assuming the impurity potential as point-like, we can calculate the
 Green's function  $G_{\sigma\sigma^{\prime}}(k,\omega,k^{\prime}.\omega)$
 using the T-matrix approximation,
\begin{equation}
 G_{\sigma\sigma^{\prime}}(k,\omega,k^{\prime},\omega) = G^{0}_{\sigma\sigma^{\prime}}\delta_{k k^{\prime}} + \sum_{\sigma^{\prime\prime}} G^{0}_{\sigma\sigma^{\prime\prime}}(k,\omega)\Gamma_{\sigma^{\prime\prime}\sigma^{\prime\prime}}G^{0}_{\sigma^{\prime\prime}\sigma^{\prime}}(k^{\prime},\omega),
\end{equation}
\begin{equation}
 \Gamma_{\sigma\sigma^{\prime}} = V^{\mbox{imp}}_{\sigma\sigma^{\prime}} + V^{\mbox{imp}}_{\sigma\sigma^{\prime\prime}}\sum_{k^{\prime}} G^{0}_{\sigma^{\prime\prime}\sigma{\prime\prime\prime}}(k^{\prime},\omega)\Gamma_{\sigma^{\prime\prime\prime}\sigma^{\prime}},
\end{equation}
 and 
\begin{equation}
 V^{\mbox{imp}}_{\sigma\sigma^{\prime}} = \left(
\matrix{
 0  & V_{\mbox{imp}} e^{i \phi} \cr
 V_{\mbox{imp}} e^{-i \phi} & 0 \cr  
}
\right)\delta_{\sigma\sigma^{\prime}}.
\end{equation}

After some calculation  with the assumption  
$\epsilon^a(k)= - \epsilon^b(k)$ and constant density of states, we get
\begin{equation}
 \Gamma_{\sigma\sigma^{\prime}} = V^{\mbox{imp}}_{\sigma\sigma^{\prime}} ( 1- \sum_{k}G^0(k)V^{\mbox{imp}})^{-1},
\end{equation} 
and 
\begin{equation}
 \sum_k G^0(k)V^{\mbox{imp}} = \delta_{\sigma\sigma^{\prime}} \frac{\pi N_F V_{\mbox{imp}}}{\sqrt{\Delta_\sigma^2 - (\omega + \mu)^2}}\left(
\matrix{
 \Delta_\sigma e^{-i\phi} &  ( -\omega -\mu) e^{i \phi} \cr
 (-\omega-\mu) e^{-i \phi}  & \Delta_\sigma e^{i \phi} \cr 
}
\right),
\end{equation}
 here $N_F$ is the density of states at Fermi energy.

 The
determinant of matrix $ D = 1-\sum_k
G^0_{\sigma\sigma}(k)V^{\mbox{imp}}$
is given as
\begin{eqnarray}
\mbox{det} D& = &(1-\frac{ \tilde{V}\Delta_\sigma e^{-i \phi}}{\sqrt{\Delta^2_\sigma - (\omega+\mu)^2}})(1-\frac{\tilde{V} \Delta_\sigma e ^{i \phi}}{\sqrt{\Delta^2_\sigma - (\omega +\mu)^2}}) -\frac{\tilde{V}^2 (\omega+ \mu)^2}{\Delta^2_\sigma - (\omega+\mu)^2}\nonumber\\
 & = & 1+\tilde{V}^2 - 2 \frac{\tilde{V} \Delta_\sigma\cos \phi} {\sqrt{\Delta^2_\sigma - (\omega+\mu)^2}},
\end{eqnarray}
where $\tilde{V} = \pi N_F V_{\mbox{imp}}$. From this expression, we find 
that when $\mbox{Re}(\tilde{V}\Delta\cos\phi)$ is positive $D^{-1}$ has  poles at $\omega= -\mu \pm \sqrt{(1-\frac{4\tilde{V}^2\cos^2
\phi}{(1 + \tilde{V}^2)^2})}\Delta_\sigma$ and two  bound states appear.
On the other hand, when $\mbox{Re}
(\tilde{V}\Delta \cos \phi) $ is negative,
there exists no bound state. We conclude that the phase of the 
order parameter controls  the energy and existence of bound states. If
 we set $\Delta$ real and positive, there exist bound states at $\omega= -\mu \pm
\frac{(1-\tilde{V}^2)}{(1 + \tilde{V}^2)}\Delta_\sigma$ when
$\phi=0$. Here we note that this bound state exists even if
$V_{\mbox{imp}}$ is very weak.
As $\phi$ changes from 0 to $\pi/2$, the energies  of bound
states approach to $\omega= -\mu \pm \Delta$ and at $\phi = \pi/2$ the
bound states are touch the edges of  the continuum.

 It should be noticed there exists a great difference in the effect
 of impurities between the singlet excitonic insulator  and the triplet
 one. In the singlet excitonic insulator,
 $\Delta_{\uparrow}=\Delta_{\downarrow}$ and both up-spin
 and down-spin bound states have the same energy. On the other hand,
 in the triplet excitonic insulator, $\Delta_\uparrow = -
 \Delta_\downarrow$ and 
 the bound states of up-spin and down-spin
 electrons cannot
 exist at the same  time. Therefore spins of the electrons doped into
 the triplet excitonic insulator align ferromagnetically. For
 $\phi=\pi/2$,
 marginal bound states exist for both spin states and excitonic
 insulator remains paramagnetic.
 
\section{Reduction of Energy gap by impurities}
\label{sec:gap-reduction}

 In the previous section, we studied the formation of  a  the bound
 state around an
 impurity. We found that a bound state appears only when
 $\mbox{Re}(\tilde{V}\Delta\cos\phi)$ is positive. However, the impurity will also influence
  the continuum state. For example, it is well known that magnetic
  impurities in a superconductor reduce the energy
 gap. Zittartz discussed the effect of intraband scattering by
 impurities  in excitonic insulator and found that it  reduces  the energy gap
 in exactly the same way as  magnetic impurities do in a superconductor\cite{Zittartz}. In this section, we
 consider the reduction of energy gap by interband scattering term.

   To study the suppression of the energy gap, we
 use the same technique as Abrikosov and Gor'kov\cite{Abrikosov1} used
 for a superconductor.
 For simplicity we again assume that the order parameter is diagonal in
 spin space.
 We introduce  the
  renormalized Green's function $\tilde{G}$
\begin{equation}
 \tilde{G}_{\sigma\sigma^{\prime}}(k,\omega)=\left(
\matrix{
 \tilde{\omega} - \epsilon^a_k + \mu & -\tilde{\Delta}_\sigma \cr
 -\tilde{\Delta}_\sigma^{*} & \tilde{\omega}-\epsilon^b_k + \mu \cr
}
\right)^{-1}\delta_{\sigma\sigma^{\prime}} .
\end{equation}
 Here $\tilde{\omega}$ and $\tilde{\Delta}$ are the renormalized frequency
 and order parameter respectively, and  we  recovered  the translational invariance   of Green's
 function of continuum states by averaging over the impurity positions.
 
 Using the Born approximation, we get 
\begin{equation}
 \tilde{G}_{\sigma\sigma^{\prime}}(k,\omega)^{-1} = G^0_{\sigma\sigma^{\prime}}(k,\omega)^{-1} - \Sigma_{\sigma\sigma^{\prime}}(k,\omega)\label{renomarlizedGreen}
\end{equation}
and 
\begin{equation}
 \Sigma_{\sigma\sigma^{\prime}}(k,\omega) = n_{\mbox{imp}}\int \frac{d^3 k^{\prime}}{(2\pi)^3} V^{\mbox{imp}}_{\sigma\sigma^{\prime\prime}}\tilde{G}_{\sigma^{\prime\prime}\sigma^{\prime\prime\prime}}(k^{\prime},\omega) V^{\mbox{imp}}_{\sigma^{\prime\prime\prime}\sigma^{\prime}}.
\end{equation}
Integrating over $k^{\prime}$ we obtain 
\begin{equation}
  \Sigma_{\sigma\sigma^{\prime}}(k,\omega) =\frac{V^2_{\mbox{imp}} n_{\mbox{imp}} \pi N_F}{\sqrt{|\tilde{\Delta}_{\sigma}|^2 - (\tilde{\omega}+\mu)^2}}\left(
\matrix{
\tilde{\omega}+\mu  & -\tilde{\Delta}_{\sigma}^{*}e^{2 i \phi} \cr 
\tilde{\Delta}_{\sigma}e^{-2 i \phi} & -\tilde{\omega}+\mu \cr
}
\right)\delta_{\sigma\sigma^{\prime}},
\end{equation}
where we  assumed $\epsilon^a_k = -\epsilon^b_k$  and constant density
of states as before. After some calculations we get the following equations for $\tilde{\omega}$ and $\tilde{\Delta}_{\sigma}$ 
\begin{equation}
 \tilde{\omega}= \omega + \frac{1}{\tau}\frac{\tilde{\omega}+\mu}{\sqrt{|\tilde{\Delta}_{\sigma}|^2 - (\tilde{\omega}+\mu)^2}}\label{omegatilde}
\end{equation}
\begin{equation}
 \tilde{\Delta}_{\sigma}=\Delta_{\sigma} - \frac{1}{\tau}\frac{\tilde{\Delta}_{\sigma}^{*}e^{-2 i \phi}}{\sqrt{|\tilde{\Delta}_{\sigma}|^2 - (\tilde{\omega}+\mu)^2}}.\label{deltatilde}
\end{equation}
 Here $\tau=1/n_{\mbox{imp}} N_F V^2_{\mbox{imp}}$. If $\phi=0$, these
 equations are the same as those for  magnetic impurities in
 a superconductor. Therefore if $\phi=0$, the energy gap is strongly
 suppressed by impurities. As $\phi$ develops from 0, the energy
 gap increases and at $\phi= \pi/2$, the energy gap becomes almost as same as
 that of pure excitonic insulator, because at $\phi=\pi/2$ the equations
 (\ref{omegatilde}) and (\ref{deltatilde}) have the same form as those
 for  non-magnetic impurities in a superconductor.
 
\begin{center}
\begin{figure}
 \epsfig{file = 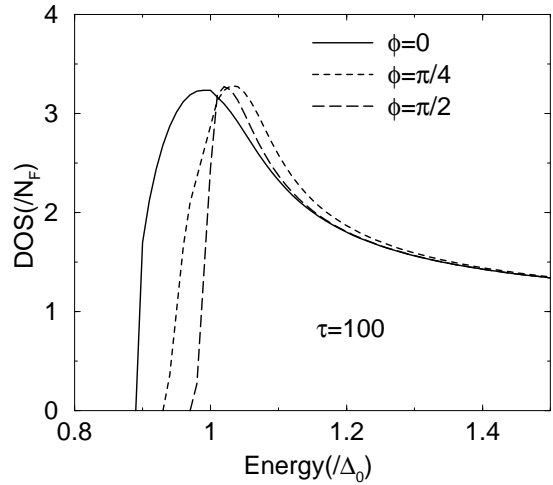, width =0.4\textwidth}
 \caption {phase dependence of DOS when $\tau$=100}
 \label{dosfigure}
\end{figure} 
\end{center}

 The density of states of electrons $N(\omega)$ is given by 
\begin{eqnarray}
N(\omega) &=& -\frac{1}{2\pi}\mbox{Im}\int\mbox{Tr}G(k,\omega)\frac{d^3 k}{(2\pi^3)} \nonumber\\
 & = &- N_F\mbox{Im}\sum_{\sigma}(\frac{\tilde{\omega }+ \mu}{|\tilde{\Delta}_{\sigma}|^2 - (\tilde{\omega}+\mu)^2})
\end{eqnarray}
 For some typical values of $\phi$, we plot the density of states in
 Fig. \ref{dosfigure}. From this figure, we find that the reduction of
 the energy gap is maximum at $\phi=0$. This is  consistent with the above discussion. 

\section{Numerical Solution of BdG equation}
\label{sec:bdg}

From the analysis in  the previous sections, we found  that the phase of the order parameter determines the existence of
the bound
state as well as  the reduction of the energy gap. These two
effects  are
not independent. For example, for  the finite concentration of impurities, the
 bound states will form an impurity band. On the other hand, if the
 energy gap of the continuum is strongly reduced, the energy of the
 bound state present for a single impurity model may be absorbed into the
 continuum for a  finite impurity concentration.  To
 study such  possibilities, in this section we solve the Bogoliubov-de 
Gennes equations for excitonic insulator numerically.
We confine ourselves to the  triplet case, because in the triplet
excitonic insulators impurity may invoke
magnetic moment as discussed  in section \ref{sec:bound-state}. We also
show that incomplete ferromagnetism can appear.

We start from the following Hamiltonian
\begin{eqnarray}
 H_{\mbox{BdG}}& = & \sum_{\langle i,j \rangle,\sigma}( -t^a a^{\dagger}_{i\sigma} a_{j\sigma} - t^b b^{\dagger}_{i\sigma}b_{j\sigma}) - \sum_{i,\sigma,\sigma^{\prime}} (\Delta_{i,\sigma\sigma^{\prime}}b^{\dagger}_{i\sigma^{\prime}}a_{i\sigma}+ c.c.)\nonumber\\
& &  + V_{\mbox{imp}}\sum ( b^{\dagger}_{i\sigma}a_{i\sigma} + c.c.) - E_a\sum_{i,\sigma}a^{\dagger}_{i\sigma}a_{i\sigma} -E_b\sum_{i,\sigma}b^{\dagger}_{i\sigma}b_{i\sigma}.\label{bdgHamiltonian}
  \end{eqnarray}
The first term is the kinetic term of electrons. Here we assume
two-dimensional square lattice and the summation is carried out with
nearest neighbor $\langle i,j \rangle$. In the theory of the excitonic
 insulator, it is often assumed that $\epsilon^a_k = -\epsilon^b_k$ , so
 we take $t^a = -t^b=1$. In this case, $\epsilon_k^a = -2 (\cos k_x +
 \cos k_y) -E_a$ and $\epsilon^b_k = 2 (\cos k_x + \cos k_y) -E_b$. The second term is the interaction term. The
 order parameter $\Delta_{i,\sigma\sigma^{\prime}}$ is determined from
 the Hamiltonian (\ref{cleanHamiltonian}) as 

\begin{equation}
 {\Delta}_{i\sigma\sigma^{\prime}}=V\langle a^{\dagger}_{i\sigma}b_{i\sigma^{\prime}}\rangle -U \delta_{\sigma\sigma^{\prime}}\sum_{\sigma^{\prime\prime}}\langle a^{\dagger}_{i\sigma^{\prime\prime}}b_{i\sigma^{\prime\prime}}\rangle.
\label{op-equation-realspace}
\end{equation}
While for $U=0$ triplet state and
 singlet state are degenerate, positive $U$ suppresses
 singlet one. In the following numerical calculation we
 take $U$ large enough to suppress the singlet state completely.
 The third term describes the impurity scattering and
the sum is carried out on the impurity site. The last two terms are the
energy gap term, which determines the overlap of two bands. In the
 following, we set $E_a=-E_b$. Under this assumption, the Fermi surfaces of
 two electron bands coincide when we dope no electron. The size
 of the Fermi surfaces depends on $E_a$. For 
 $E_a <  -4t$, $\epsilon^a_k < \epsilon^b_k$ for  any $k$, and the system
 is a  band insulator. On increasing $E^a$, the Fermi surface emerges
 near $k=(\pi,\pi)$ and the system becomes a semi-metal. When  $E_a > 4t$
 it becomes a band insulator again, because $\epsilon^b_k <
 \epsilon^a_k$.

 We solve equation (\ref{bdgHamiltonian}) and
 (\ref{op-equation-realspace}) numerically keeping the number of electrons constant.  
First, we
 diagonalize the equation(\ref{bdgHamiltonian}) and calculate the
 eigenenergy and eigenstates. Next we take the eigenstates with  small
 eigenvalues, up to the number of electrons. The order parameter is
 calculated by equation (\ref{op-equation-realspace}). We repeat these procedures
 iteratively until the solution converges. We change the phase $\phi$ by
 choosing the initial values of $\Delta$, and check the phase after
 convergence. 

 We set parameters $U=1.0$, $V=2.0$, $V_{\mbox{imp}}=1.0$ and $E_a =-E_b
 = 0$. The calculation is carried out on $17\times 17$ lattices.

\begin{figure}
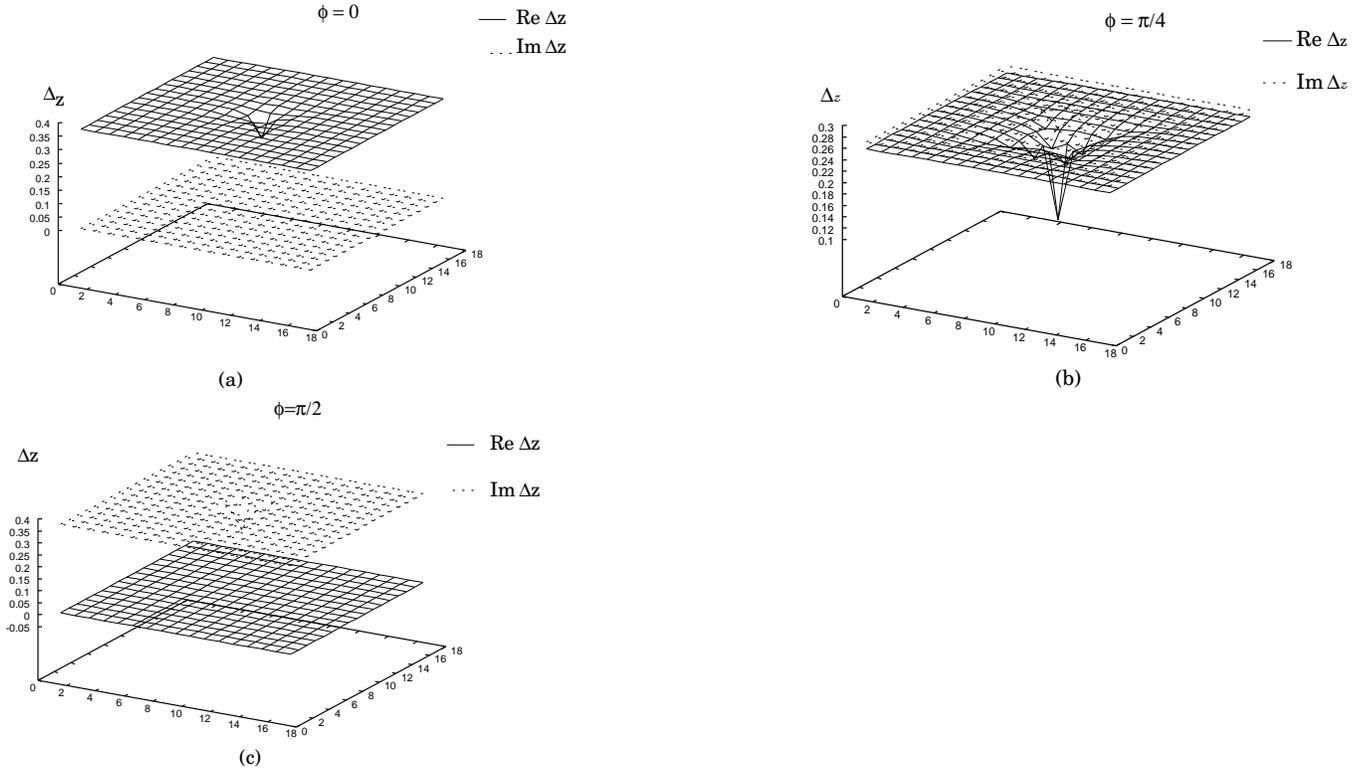

 \epsfig{file = op-phase0.eps, width =0.4\textwidth}
 \epsfig{file = op-phase4.eps,width = 0.4\textwidth}
 \epsfig{file = op-phase2.eps, width= 0.4\textwidth}
 \caption {Real and imaginary part of triplet order parameter when $\phi = 0$,$\phi=\pi/4$ and $\phi = \pi/2$.}
 \label{op-phase}
\end{figure}

 First we consider the case when the impurities are introduced into the
 non-doped excitonic insulator. In Fig. {\ref{op-phase}} we present the
 triplet order parameter $\Delta_z$ axis when one impurity is
 introduced, setting the initial condition $\phi = 0, \pi/4$ and
 $\pi/2$. The order parameter is  reduced 
 at the impurity site. We note that in Fig. {\ref{op-phase}}(b) $\phi$
 differs from $\pi/4$. This is due to the fact that after iteration
 $\phi$ deviates from the initial values.
 However, from Fig. {\ref{op-phase}} we find that the difference is
 small. 

\begin{center}
\begin{figure}
 \epsfig{file = 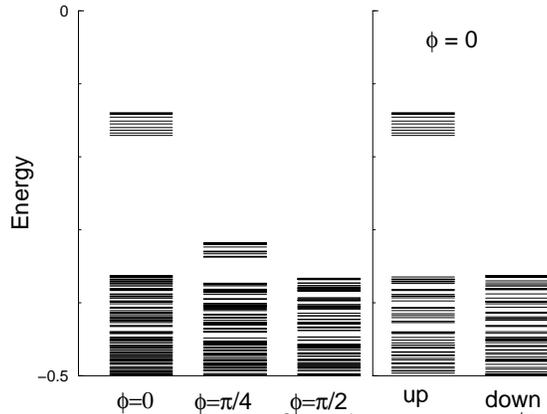, width =0.4\textwidth}
 \caption {phase dependence of energy spectra near $\epsilon=0$. phase $\phi$ is set to
 $0$, $\pi/4$ and $\pi/2$. The right figure shows the spectrum of
 up-spin and down-spin electrons separately, when $\phi=0$.} 
 \label{spectre}
\end{figure} 
\end{center}

 To see the reduction of energy gap, we need finite concentration of
 impurities.  In Fig. \ref{spectre} we plot the single particle spectrum near
 $\epsilon = 0$ when 10 impurities are introduced periodically. We find
 that the bound  states appear
 around $E=0.2$ when $\phi=0$. It should be noted that the all
 electrons in the bound state has up-spin, as shown in the right of 
 the figure. The number of localized eigenstates with negative energy is
 the same as that of impurities.   The energy of the bound states become
 large as $\phi$ changes from $0$ to $\pi/2$, and vanish at $\pi/2$, in
 consistent with our discussion.  We also note that at $\phi = \pi/4$
 the energy gap is larger than the case when $\phi = 0$. This is
 consistent with the result of section \ref{sec:gap-reduction}. We note
 that at $\phi=\pi/2$ the gap seems to be smaller than the case when
 $\phi =\pi/4$. At $\phi = \pi/2$, the bound states touch the edge of
 the continuum and hybridize with the continuum. The energy of
 eigenstates is lowered by this hybridization.  

\begin{figure}
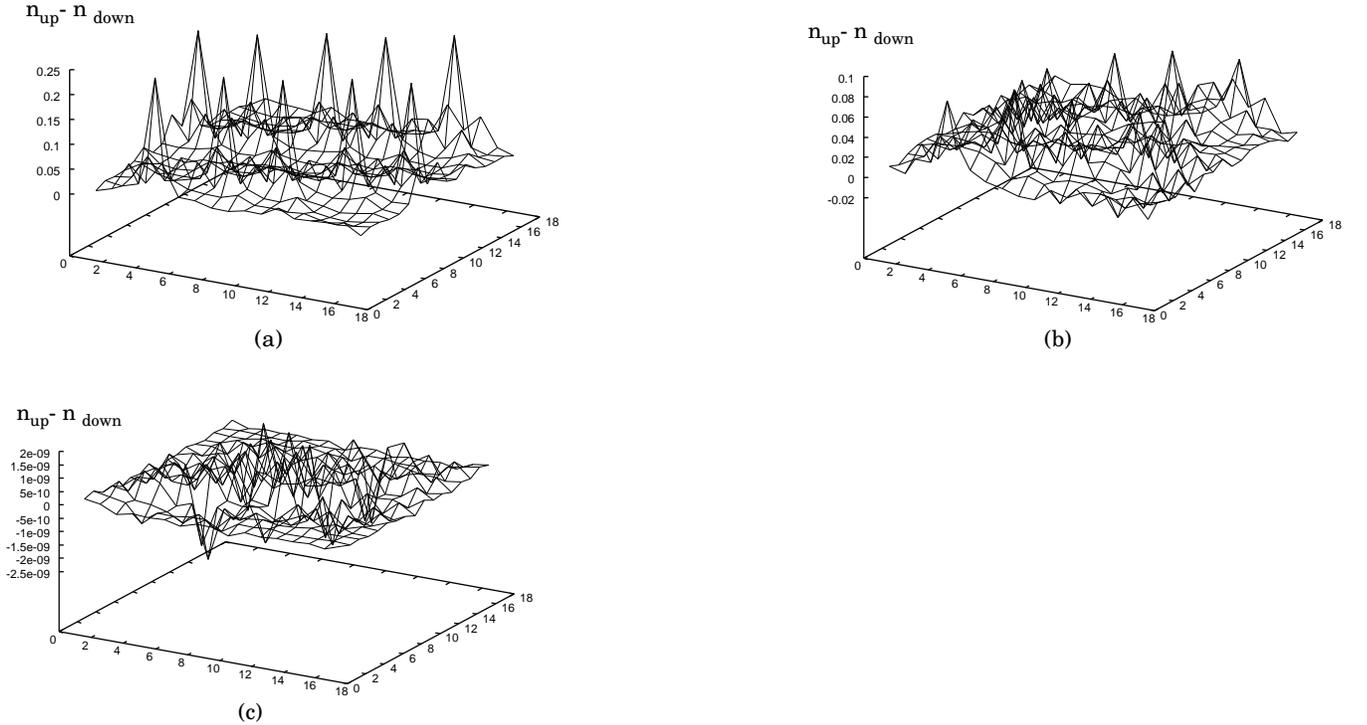

 \epsfig{file = mag-10-10-0.eps,width=0.4\textwidth}
 \epsfig{file = mag-10-10-716.eps,width=0.4\textwidth}
 \epsfig{file = mag-10-10-2.eps,width=0.4\textwidth}
 \caption {Spatial pattern of magnetization when (a) $\phi=0$, (b)
  $\phi=7\pi/16$ and (c) $\phi = \pi/2$, when the number of
  electrons and the number of impurities are both 10.}
 \label{magnetic-moment}
 \end{figure}

 Now we turn to the case when electrons are doped.
When doped into a  triplet excitonic insulator, electrons
will align ferromagnetically  if $\phi \neq \pi/2$. In the case of
La$_x$Ca$_{1-x}$B$_6$, La acts as both the impurity and electron
donor. Therefore we consider the case when the number of impurities and
doped electrons are the same. In 
 Fig.\ref{magnetic-moment} we
 plot $n_{i,\uparrow}-n_{i,\downarrow}$ when 10 electrons and impurities
 are doped for $\phi=0,7\pi/16,\pi/2$. When $\phi=0$ we can see that
 localized magnetic
 moment appears around the impurity. Total magnetic moment amounts to
 10, which means that all doped electrons have spin up. On
 the other hand, when $\phi=\pi/2$ no magnetic moment appears. Therefore
 we can conclude when $\phi=\pi/2$ the doped triplet excitonic insulator
 is paramagnetic. These
 results are also consistent with our discussion. In the middle of these
 two limits, there may exists incomplete ferromagnetism. In
 Fig.\ref{magnetic-moment}(b), we show the same plot for 
 $\phi=7\pi/16$. We can see that magnetic moment become small, being 6
 in total.  This incomplete ferromagnetism is considered as
 follows. If the concentration of impurities is finite, associated
 impurity bound states form an impurity band. Since the bound state
 energy approaches $\Delta$ as $\phi\rightarrow\pi/2$, this impurity
 band has a finite overlap with the continuum state for $\phi$
 sufficient close to $\pi/2$. In this case ferromagnetic moment due to
 doped electrons will be reduced.

\section{Conclusion and Discussion}
\label{sec:conclusion}

 On the basis of the results of the preceding sections, we first
 discuss  the weak ferromagnetism in doped hexaborides. As
 discussed in this paper,  doped triplet excitonic insulator with
 impurities shows  ferromagnetism if phase $\phi$
 is between $-\pi/2$ and $\pi/2$. The incomplete ferromagnetism can also
 occur when the energy gap becomes smaller than the bound state energy.
 Here we discuss the possibility that the ferromagnetism of La-doped
 CaB$_6$ is caused by impurities. We assume that CaB$_6$ is the  triplet
 excitonic insulator.
 When we dope La, La acts as impurity and cause interband scattering.
 As discussed in this paper, in the presence of interband scattering
 up-spin and down-spin does not compensate
 because of the emergence of localized moment if $\phi \neq
 \pi/2$. Therefore ferromagnetism emerges, which become incomplete when
 the reduced energy gap is smaller than the highest energy of impurity band.
 Here we discuss the possibility of the impurity-induced ferromagnetism
 by this mechanism.

 First, we notice the ferromagnetism emerges only when excitonic order
  is spin-triplet.  It is well known that direct interaction
 term suppresses the singlet order, while exchange  interaction stabilize both
 singlet and triplet order by an equal amount. Therefore only triplet order
 parameter will be realized when direct interaction is large
 enough. It seems natural to assume that the exchange interaction is  the same
 order as the direct interaction. As shown in sec.\ref{sec:bdg}, triplet
 excitonic insulator is realized when $U$ and $V$ are the same order.

 Second, the phase of order
 parameter must not be $\pi/2$, in other words, the minimum of the total
 energy is achieved when $\phi\neq\pi/2$. In our discussion, we do not
 discuss the value of $\phi$ which minimize the total energy. In our
 numerical  calculation the change of energy by phase is very small.  As
 noted in sec.{\ref{sec:breaking-symmetry}} in  real materials there
 exist some other interactions which is not discussed here, such as
 interband pair scattering. These interactions  also depend on the phase.
 Because these interactions exist even without impurities, we think that
 the stable phase $\phi$
 will be  determined by such term.Further study will be needed to determine the stable
 $\phi$, however, there exists no reason to restrict $\phi=\pi/2$.

 Third, the minimum of the energy gap must be smaller than the energy of
 the impurity band. In our calculation it means that  $\phi$ is close to
 $\pi/2$. However, it should be noticed that there exists other origins
 of the  reduction of the energy gap. For example, intraband scattering
 by impurities also 
decreases the energy gap, as discussed by Zittartz\cite{Zittartz}.
In this paper we assume La acts as the interband scattering potential.
However, there exists other impurities or disorder in
La$_x$Ca$_{1-x}$B$_6$. These impurities and disorder will also acts as
impurity potential. If these scattering is large enough, the energy gap
will be reduced such that the energy of the bound states is larger than
the gap. We conclude that the
minimum of the gap can be smaller than the energy of the bound states.

 One may think  that our model is too simplified to describe the La-doped
 hexaborides. In the following we briefly comment on the physics
 neglected in our model. 

 First, we only consider one pair of Fermi surfaces, 
 which shows completely nesting without doping. However,
 the band calculation of CaB$_6$ shows that there exist three pairs of
 imcompletely nested Fermi surfaces. 
 Balents and Varma\cite{Varma1} discussed that in CaB$_6$
 we must consider the possibility of the co-existence of gapped and
 ungapped Fermi surfaces. In this case, though our theory should be
 modified, ferromagnetism will occure. Our theory is based on the fact
 that Green's function  depends on the phase. Even if gapless
 Fermi surfaces  exist, Green's function
 $G_{\sigma,\sigma^{\prime}}(k,\omega,k^{\prime},\omega)$ will show the
 phase dependence. Therefore the occurence of ferromagnetism will be
 suggested in this case.
 On the other hand, the formation of LOFF state discussed by Barzykin and
 Gorkov\cite{Gorkov1} gives severe damage to our theory. If LOFF state
 appears, order
 parameter $\Delta$ shows spatial modulation,  such as  $\Delta(r)=\Delta
 \exp({iqr})$. In this case, total magnetic moment will diappear,
 because the direction of local magnetic moment will be random. 

 Another important point we have neglected   is the symmetry of the
 energy band. Band calculation shows that the conduction band and the valence
 band have different symmetries,  $X^{\prime}_3$ and $X_3$. Though a simple
 substitution of Ca by La  at a single site  will reduce the symmetry,
 the rotational symmetry and parity around the impurity survive. This
 leads to the conclusion that interband scattering by La is
 forbidden by the symmetry. Therefore for our theory we must assume the
 symmetry is
 lowerd at La site. For example, if the impurity locates at less symmetric
 position, these symmetries  break down and intraband scattering is
 permitted.

 We also note that the disorder in the lattice
 will also cause the intraband scattering. Though in  the above
 discussion we have assumed that only La acts as
 interband scattering potential, interband scattering by disorder will
 also happen. In this case our theory does not
 modified, except that the number of impurities is different from  that
 of doped electrons. 

 The readers will  also notice that intraband scattering is neglected in
 our model.
 However  we can show  that the bound state can survive even if
 intraband scattering
 exists.When one impurity causes both interband and intraband 
scattering, the energy of localized state is determined by
\begin{equation}
 (1+\tilde{V}_{\mbox{imp}}^2-\tilde{V}^{\prime 2}_{\mbox{imp}})(\Delta^2-(\omega+\mu)^2) + 2 \tilde{V}^{\prime}_{\mbox{imp}}(\omega+\mu) - 2 \tilde{V}_{\mbox{imp}}\Delta \cos \phi = 0 
\label{inter-intra}
\end{equation}
where $\tilde{V}^{\prime}_{\mbox{imp}} = \pi N_F V^{\prime}_{\mbox{imp}}$ and
$V^{\prime}_{\mbox{imp}}$ is the strength of intraband scattering. If  
$|\tilde{V}^{\prime}_{\mbox{imp}}| < \tilde{V}_{\mbox{imp}} \cos \phi$, equation
(\ref{inter-intra}) has two bound states when  $\tilde{V}_{\mbox{imp}}\Delta\cos\phi
>0$ . In this case the discussion above is not qualitatively modified.
On the other hand, $|\tilde{V}^{\prime}_{\mbox{imp}}| > \tilde{V}_{\mbox{imp}} \cos\phi$, equation
(\ref{inter-intra}) has one bound state, independent of  the sign of
  $\tilde{V}\Delta\cos\phi$. However here we note the energy of bound
  state differs between positive and negative $\Delta$. Therefore we
  conclude the ferromagnetism will also survive, though some modification
  will be needed.
 
 So far  there are no direct evidence of this impurity-induced
 incomplete ferromagnetism in the hexaborides.  However we note that we can account for
  the experimental inconsistencies  on the value of  the saturation moment of  La$_x$Ca$_{1-x}$B$_6$ by
 this scenario. For example, in the
 first paper of Fisk {\it et al.} the saturation moment
 is about 0.07$\mu_B$ per electron when $x\sim 0.005$, while Terashima {\it et al.} reports
 that the saturation moment is about 0.25 $\mu_B$ with same doping,
 more than three times
 larger  than that of Fisk\cite{Terashima}. We can explain this
 discrepancy by the difference of the density of impurities. More
 careful experiments on the magnetism will be needed to test the present
 theory. The measurement of the local magnetic moment will give much
 information about this.

 In summary, we have investigated the effect of the impurity in the  excitonic
 insulator. The effect  depends on the phase of the order parameter. 
 The bound state appears when $\mbox{Re} \Delta$ is positive, and the
 energy of
 the bound state depends on the phase of the  order parameter. On the other
 hand, the reduction of the energy gap of  continuum state becomes maximum
 when $\mbox{Im} \Delta=0$. Numerical solution of BdG equation
 supports these discussion. Based on these results, we proposed a new
 mechanism of incomplete ferromagnetism and applied  it to the
 hexaborides. 
Our
 theory qualitatively explains  experiments, though further study will
 be needed.

\acknowledgements
 The author acknowledges  K. Miyake, H. Kohno and  S. Maebashi for fruitful
 discussion. The numerical computation is partially carried out by the Yukawa
 Institute Computer Facility. This work is financially supported by
 Research Fellowships of the Japan Society for the Promotion of Science
 for Young Scientists.

\end{document}